\documentclass[aip,reprint,nofootinbib]{revtex4-2}

\usepackage[utf8]{inputenc}
\usepackage[T1]{fontenc}
\usepackage{textcomp}
\usepackage{graphicx}
\usepackage{amsmath}
\usepackage{mathtools}
\usepackage{color}
\usepackage[colorlinks=True]{hyperref}
\usepackage{titlesec}
\usepackage{amsfonts}
\usepackage{etoolbox}
\usepackage{physics}
\usepackage{stmaryrd}

\graphicspath{ {./Figures/revision} }

\draft 
\newcommand{\rosso}[1]{{\color{black} #1}}
\newcommand{\blu}[1]{{\color{black} #1}}

\renewcommand{\rosso}[1]{{\color{black} #1}}

\titlespacing*{\section} {0pt}{4ex}{0.5ex}

\makeatletter
\def\@email#1#2{%
 \endgroup
 \patchcmd{\titleblock@produce}
  {\frontmatter@RRAPformat}
  {\frontmatter@RRAPformat{\produce@RRAP{*#1\href{mailto:#2}{#2}}}\frontmatter@RRAPformat}
  {}{}
}%
\makeatother

\makeatletter
\newcommand{\printfnsymbol}[1]{%
  \textsuperscript{\@fnsymbol{#1})}%
}
\makeatother

\begin{document}


\title{Ultrafast ($\approx$10 GHz) mid-IR modulator based on ultra-fast electrical switching of the light-matter coupling} 



\author{Mario Malerba}
\altaffiliation{Now with Istituto Nazionale di Ricerca Metrologica (INRIM), Strada delle Cacce 91, 10135 Torino, Italy}

\author{Stefano Pirotta}
\thanks{M.Malerba and S.Pirotta contributed equally to this work.}
\author{Guy Aubin}
\author{L. Lucia}
\author{M. Jeannin}
\author{J-M. Manceau}
\author{A. Bousseksou}
\affiliation{Centre de Nanosciences et de Nanotechnologies, CNRS UMR 9001, Université Paris Saclay, 10 Boulevard Thomas Gobert, 91120 Palaiseau, France}

\author{Q. Lin}
\author{J-F Lampin}
\author{E. Peytavit}
\author{S. Barbieri}
\affiliation{Institute of Electronics, Microelectronics and Nanotechnology, CNRS, Univ. Lille, Univ. Polytechnique Hauts-de-France, UMR 8520, F-59000 Lille, France
}

\author{L. Li}
\author{A.G. Davies}
\author{E.H. Linfield}
\affiliation{School of Electronic and Electrical Engineering, University of Leeds, Leeds LS2 9JT, United Kingdom}

\author{Raffaele Colombelli}
\email[Authors to whom correspondence should be addressed:]{raffaele.colombelli@universite-paris-saclay.fr and  m.malerba@inrim.it}
\affiliation{Centre de Nanosciences et de Nanotechnologies, CNRS UMR 9001, Université Paris Saclay, 10 Boulevard Thomas Gobert, 91120 Palaiseau, France}

\date{\today}

\begin{abstract}
We demonstrate a free-space amplitude modulator for mid-infrared radiation 
($\lambda$ $\approx$  9.6 $\mu$m) that operates at room temperature up to 
at least 20 GHz (above the -3dB cutoff frequency measured at 8.2 GHz). 
The device relies on the ultra-fast transition between weak and strong-coupling regimes induced by the variation of the applied bias voltage. 
Such transition induces a modulation of the device reflectivity.
It is made of a semiconductor heterostructure enclosed in a judiciously designed array of metal–metal optical resonators, that - all-together - behave as an electrically tunable surface. At negative bias, it operates in the weak light-matter coupling regime. Upon application of an appropriate positive bias, the quantum wells populate with electrons and the device transitions to the strong-coupling regime. 
The modulator transmission keeps linear with input RF power in the 0dBm - 9dBm range.
The increase of optical powers up to 25 mW  exhibit a weak beginning saturation a little bit below.
\end{abstract}


\maketitle 


Electrically reconfigurable surfaces are artificial components whose optical properties, in reflection/absorption, can be addressed electrically~\cite{Chen2008a,Liu2016,Ketchum2021}. In particular, surfaces whose complex reflectivity (real/ imaginary parts of the S11 parameter, in electronic scattering terms) is electrically tunable are particularly useful as amplitude or phase modulators~\cite{Park2017,Benea-Chelmus2021,Chung2023,Sherrott2017}.
In the mid-IR (3um< $\lambda$<30um), these functionalities are useful for applications such as laser amplitude/frequency stabilization\cite{Bernard1997}, coherent detection, spectroscopy and sensing, mode-locking, and optical communications\cite{Spitz2022}.
However, the ultra-fast (10–40 GHz) modulation of mid-IR radiation has been for a long time a largely under-developed functionality\cite{Berger1996,Dupont1993,Herrmann2020}. Recently, advances have been proposed, in semi-guided/guided architectures, on III-V \cite{Dely2022} and on SiGe platforms\cite{Nguyen2024}. 
In the former case an optical 10 Gbit/s free-space transmission at 9 $\mu$m has been demonstrated despite a 3 dB cutoff at 3 GHz of the full system. 
In the latter case, operation was reported up to $\approx$ 1.5 GHz.
\begin{figure}[t]
\centering
\includegraphics[width=\columnwidth]{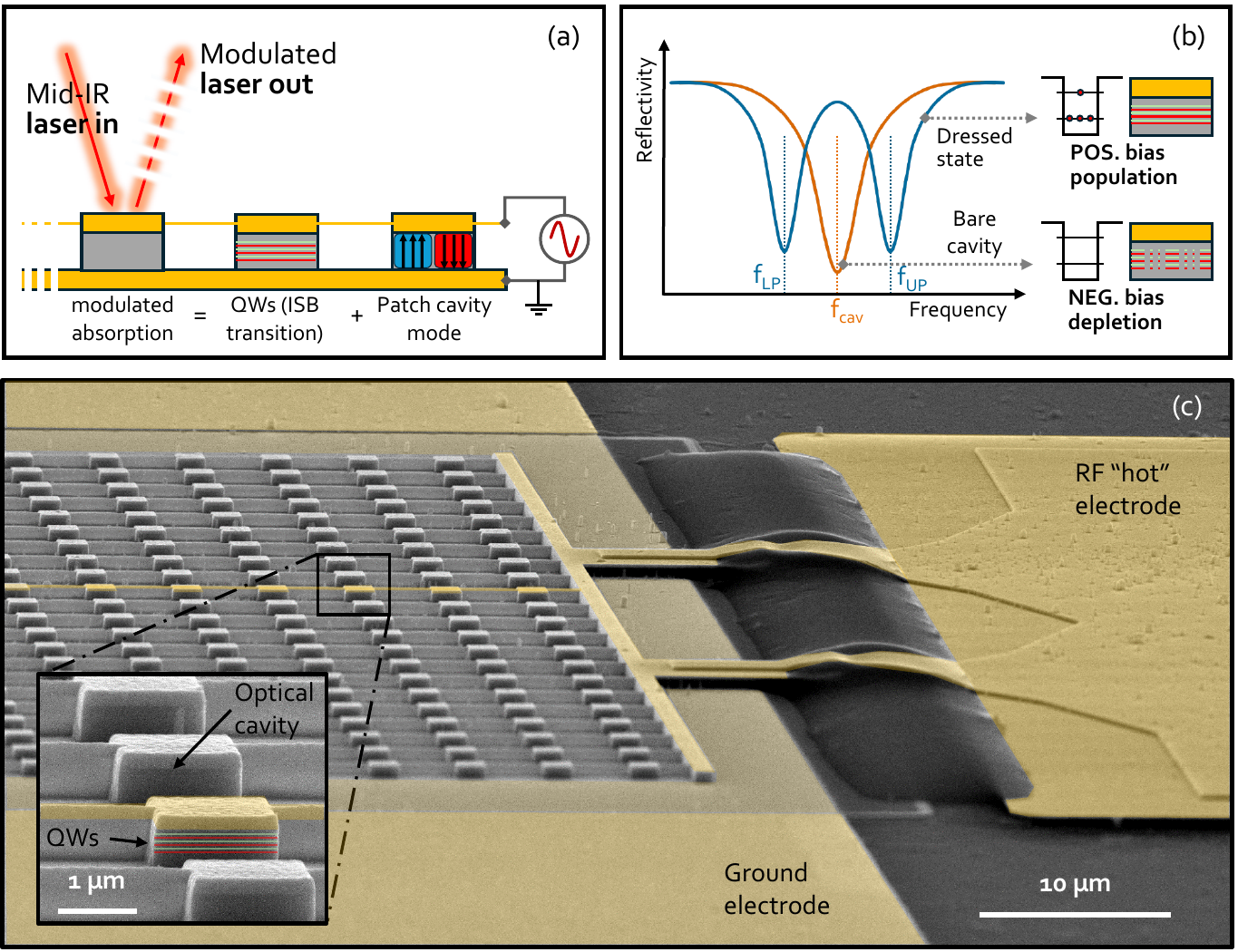}
\caption{
(a) Scheme of the modulator architecture: the active region is embedded in metal–metal cavities that are electrically connected. The amplitude of the optical reflected beam is modulated by the application of an external RF signal.
(b) Intuitive view of the modulator operating principle in an ideal configuration.
(c) SEM images of the fabricated device.}
\label{fig:1}
\end{figure}
\begin{figure*}[t]
\centering
\includegraphics[width=1.95\columnwidth]{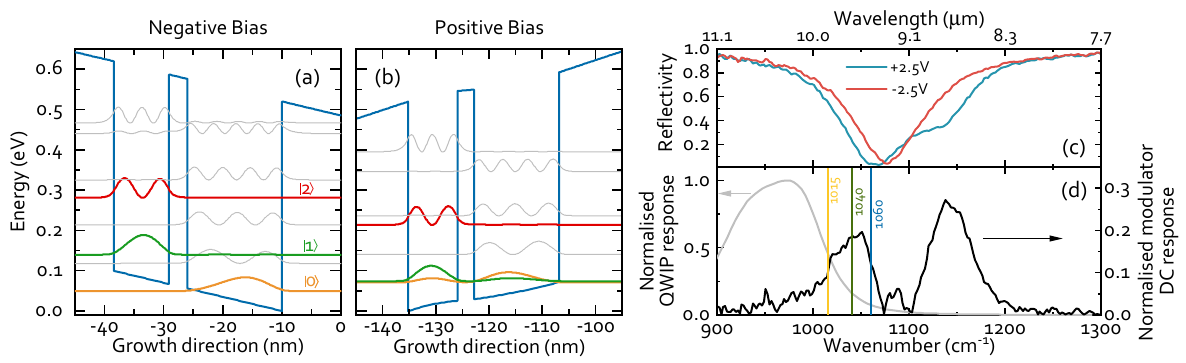}
\caption{\label{fig:2}
(a) Calculated band-structure (solid blue line), with squared modulus of the electronic wave-functions, of one period of the modulator AR at -2V applied bias. The active transition 1$\rightarrow$2 has a dipole of 2 nm, but the essentially depleted level $\ket{1}$ leads to operation in weak coupling when the structure is inserted in a patch cavity.  
(b) Calculated band-structure at +2V applied bias. Levels $\ket{0}$ and $\ket{1}$ are now aligned, electrons are delocalized and absorption takes place at the 1$\rightarrow$2 transition energy (the splitting between levels $\ket{0}$ and $\ket{1}$ is smaller than the level linewidth).   
The band-structures have been calculated with a home-written Schroedinger-Poisson solver that implements the model in Ref.~\cite{Scandolo1994}.
(c) DC reflectivity spectra of the modulator when applying -2.5V/+2.5V DC potential difference (orange and light-blue curves, respectively).
(d) Right axis: DC response of the modulator extracted from data in panel (c) following the definition in \eqref{eq:2}. Left axis: calculated fast QWIP responsivity curve (normalised) in grey (see Ref.~\cite{Lin2023})}
\end{figure*}

An alternative is to implement electrically reconfigurable surfaces for the mid-IR, where they are not as developed as in the visible/telecom spectral regions, or as in the radio-frequency (RF) domain\cite{Balanis2005}. This is the path that we have recently proposed in a proof-of-concept demonstration\cite{pirotta_fast_2021}. One of the advantages of this approach is that such devices would be immediately deployable in existing mid-IR setups that - in the vast majority - operate with free-space propagating beams.

The operating principle is described in Figs.~\ref{fig:1}(a),(b). A judiciously designed semiconductor heterostructure active region (AR), whose details are in Supplementary Material, is embedded in an array of metal–metal (MIM) optical resonators that are electrically connected so that a bias can be applied (panel (a), sketch). 
At negative bias the active QW is depleted and the AR does not absorb: only the cavity absorption peak appears in reflectivity (Fig.~\ref{fig:1}(b), orange line) at the energy $\hbar \omega_{cav}$. 
The application of a positive bias populates the QWs and activates the AR absorption at $\hbar \omega_{12}$ (see Fig.\ref{fig:2}): since the device is designed to operate in strong coupling, the reflectivity features two polaritonic resonances at energies $\hbar \omega_{LP,UP} = \hbar \omega_{cav} \pm \hbar \Omega_{Rabi}$, as sketched in Fig.~\ref{fig:1}(b), blue line.
$\hbar \omega_{LP,UP}$ are the energies of the lower (upper) polariton modes, respectively.
$\hbar \omega_{cav}$ is the energy of the cavity mode, that depends on the cavity geometry. We operate on the fundamental TM$_{10}$ cavity mode, as defined in Ref.~\cite{todorov_optical_2010}.
And $\hbar \Omega_{Rabi}$ is the Rabi energy, that gauges the strength of the light-matter coupling.
If a laser beam is shone on the device surface, and the bias is properly modulated (see sketch in Fig.~\ref{fig:1}(a)), the reflectivity change induces a corresponding amplitude modulation of the laser beam. The effect is maximum at photon energies around $\hbar \omega_{LP,UP}$ and $\hbar \omega_{cav}$. 

In this letter, we demonstrate that this approach permits to implement a free-space amplitude modulator for mid-infrared radiation 
($\lambda$ $\approx$ 9.6 $\mu$m) that operates at room temperature up to at least 20 GHz, with an optical -3 dB cutoff frequency around 10 GHz. 

The InGaAs/AlInAs semiconductor heterostructure was grown by molecular beam epitaxy lattice-matched on a low-doped InP substrate. It is composed of 8 periods of a double QW structure separated by 14-nm-thick AlInAs barriers. Each double-QW is composed of a 9.3-nm-thick active InGaAs well, separated by a 3.1-nm-thick AlInAs barrier from  a larger 16-nm-thick InGaAs QW. The latter well is delta doped in the center with Silicon to a nominal level of 1.2E12 cm$^{-2}$. The AR is preceded and capped by 40 nm and 10 nm InGaAs, respectively. The structure is inspired from a GaAs/AlGaAs design reported in Ref.~\cite{Anappara2012}, and details are reported in Supplementary Information.

At negative bias, the electronic band structure of a period of the AR is reported in Fig.~\ref{fig:2}(a). The active transition is  1$\rightarrow$2, designed at $\lambda=8.85\ \mu$m, but the active QW  (the narrower QW in the figure) is depleted, as electrons are localized in the right well. As a consequence there is no absorption: once the AR is inserted in a MIM resonator only the cavity dip appears in reflectivity.\\
This behavior persists up to 0V, where only a small fraction of the electrons populates state $\ket{1}$. As a consequence, the device operates in weak coupling and only one dip is apparent in reflectivity.\\
When a positive bias is applied, level $\ket{0}$ is progressively brought in alignment with level $\ket{1}$: 
carriers delocalize \textit{via} tunneling and the  1$\rightarrow$2 absorption increases. 
The quantity that gauges this process is the plasma frequency $\omega_{pl}=\sqrt{\frac{f_{12}n_{2D} e^2}{\epsilon_{InGaAs} \epsilon_{0} m* L_{QW}}}$, where $f_{12}$ is the oscillator strength of the active transition (the 1$\rightarrow$2 transition in Fig.~\ref{fig:2}(a)); $n_{2D}$ is
the surface electronic density difference between subbands 1 and 2; $m^*$ is the active QW effective mass (the narrower QW in Fig.~\ref{fig:2}(a), the material is InGaAs); $L_{QW}$ is the active QW width; $\epsilon_{InGaAs}$ is the dielectric constant of the QW material (InGaAs).
Once in a MIM resonator, operation takes place in  the strong coupling regime if the Rabi splitting $2 \Omega_{Rabi}= \sqrt{f_w} \omega_{pl}$ is larger than intersubband (ISB) and cavity linewidth ($f_w$ is the overlap of the active QWs with the cavity electromagnetic mode, that can be approximated as the \textit{total} thickness of active QWs divided the thickness of the active region).

The full modulators rely on MIM electrically connected patch resonators integrated with a coplanar waveguide for RF operation (see SEM image in Fig.~\ref{fig:1}(c)). 
\blu{
The details of the fabrication are  described in Supplementary Material, but
a short description is given here. 
The heterostructure is first patterned (contact optical lithography) with a 3-layered Ti-Au-Ti (5/200/20 nm) metal to implement the input coplanar RF-line (where wirebonding will take place), as well as the bottom continuous metal layer of the MIM cavities. A 35-nm-thick HfO$_2$ layer is inserted between the metallization and the AR to act as a gate. 
The sample is then epoxy-bonded to a SI-GaAs host substrate. 
The original InP substrate is removed in HCl until the InGaAs etch-stop is reached. The 435-nm-thick active region is now patterned on the bottom side, and ready for further processing on the top side. 
100 kV e-beam lithography is employed to define the patch structures and the 100-nm-wide connecting wires; Ti-Au (10/200 nm) is deposited as top contact and optical confinement layer; finally the patch cavities are defined by ICP-RIE etching using the metal itself as a self-masking element. 
The last step of the process is the implementation of the air-bridge connections between the cavities and the RF central coplanar line, and it is described in Supplementary Material. Wirebonding of the device to a PCB/SMA coplanar line completes the realization of the device.}

Different devices have been fabricated by changing the length \textit{s} of the patch cavities side (see Fig.\ref{fig:1}(a)), in the range 1.3-1.7 $\mu$m. The active region being very thin (435 nm), the system operates in the independent resonator regime: therefore  the cavity resonant fundamental frequency $\nu_{cav}$ can be tuned with \textit{s} based on the following equation~\cite{Manolatou2008d,todorov_optical_2010}:
\begin{equation}
    \frac{c}{\nu_{cav}}=\lambda_{cav}=2\ s\ n_{eff}
    \label{eq:1}
\end{equation}
$
$
A good choice of \textit{s} for a modulator is such that $\nu_{cav} \approx \nu_{ISB}$, where $\nu_{ISB}$ is the frequency of the ISB transition.
In this situation the spectrum in strong coupling shows two polaritonic dips of equal amplitude, and the modulator response is the same at both frequencies. Of course this is not a requirement and a different choice can be justified by the application.
The device reflectivities as a function of a DC applied bias have been probed with an infrared microscope connected with a Fourier transform infrared (FTIR) spectrometer. A typical result is reported in Fig.~\ref{fig:2}(c).
The DC modulator contrast can be extracted from the data as:
\begin{equation}
    {|R_{min-bias}-R_{max-bias}|}\:,
    \label{eq:2}
\end{equation}
where $R_{min-bias (max-bias)}$ is the reflectivity value at the minimum (maximum) applied bias on the modulator, respectively. From this first generation devices, we can expect a maximum modulation contrast  in the 0.15 - 0.20 range.

From the device surface (composed of all the patch cavities, the electrical connections between the cavities, and the regions where the air-bridges land) we can estimate a capacitance of 0.37 pF. This leads to  a cut-off frequency $f_{-3dB} = 1/(2\pi RC)$ of about 8 GHz, R being the 52.5 Ohm value corresponding to the access circuit, as the device impedance is very high, $\geq$ 1 M$\Omega$, and - being in parallel to the capacitance - does not affect the cut-off. We have used an equivalent circuit, albeit simplified, to the one employed in Ref.~\cite{hakl_ultrafast_2021} (See Supplementary Material for details).

\begin{figure}[b]
\centering
\includegraphics[width=0.95\columnwidth]{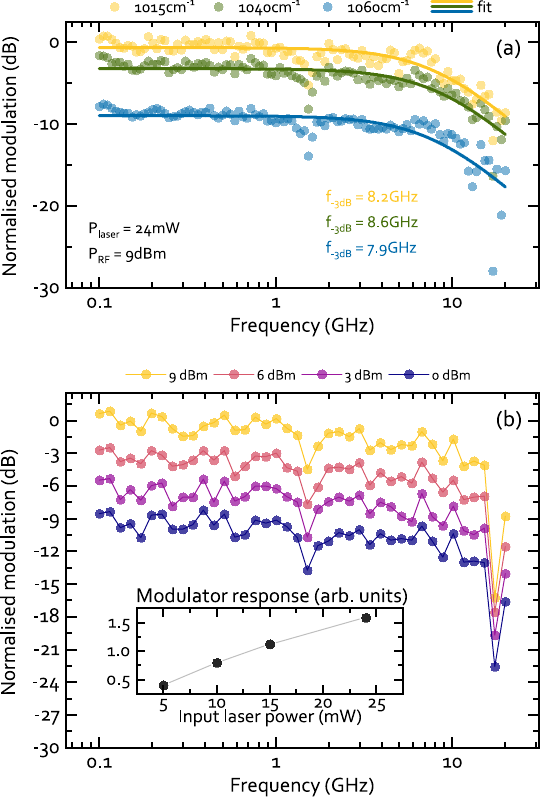}
\caption{\label{fig:3}
(a) Modulator bandwidths (filled circles) measured at different laser wavelengths (CW power 24mW):
9.85 $\mu$m (1015 cm$^{-1}$), 9.61 $\mu$m (1040 cm$^{-1}$) and 9.43 $\mu$m (1060 cm$^{-1}$). Solid lines are data best-fits with a 1$^{st}$-order low-band-pass curve: for each data set, the f$_{-3dB}$ cut-off is reported too. The dip at 1.5 GHz is a resonance in the detection chain.
(b) Modulator bandwidths for increasing RF input power: 0, 3, 6 and 9 dBm. The laser power and wavelength are 24 mW and  9.85 $\mu$m (1015 cm$^{-1}$), respectively. The modulator operation is linear with input RF power.
\textit{Inset}: modulation response as a function of the laser power. An initial onset of possible saturation is visible at 25 mW.
}
\end{figure}
We have measured the electro-optic modulation bandwidth (BW) of the modulators with the setup
described in Supplementary Material (Fig.~S1), that was also used in our previous work Ref.~\cite{pirotta_fast_2021}. In particular - for the BW measurements - we have employed a home-made fast quantum well infrared photodetector (QWIP), based on the architecture presented in Ref.~\cite{hakl_ultrafast_2021}.
Figure~\ref{fig:2}(d)(left axis) reports (grey solid line) the calculated spectral responsivity of the QWIP device we use, from Ref.~\cite{Lin2023}.
The modulator operates on the tail of the spectral responsivity curve  of this type of detectors, which however exhibits a flat bandwidth up to more than 40 GHz at room temperature. Such speed is crucial to faithfully characterize the frequency response of the modulators.

Figure~\ref{fig:3}(a) reports the modulation amplitude versus frequency when the device is driven with 9 dBm RF input power, and 24 mW incident laser CW power. 
The measurements have been acquired for different laser wavelengths (see solid vertical yellow, green and blue lines in Fig.\ref{fig:2}(d), namely: 9.85 $\mu$m (1015 cm$^{-1}$), 9.61 $\mu$m (1040 cm$^{-1}$), 9.43 $\mu$m (1060 cm$^{-1}$). The change in modulation amplitude with wavelength is mainly due to operation on the tail of the QWIP detector response curve.\\
The data (filled circles in Fig.\ref{fig:3}(a)) can be fitted with a 1$^{st}$-order low-band-pass curve\footnote{The fit function is : $R(\nu)=R_0 + 20\times\log_{10}\left(\frac{\nu_c}{\sqrt{\nu_c^2 + \nu^2}}\right)$ where $\nu_c$ is the -3dB cut-off frequency.} (solid lines), revealing that the devices operate up to 20 GHz (the limit of our spectrum analyzer). An average -3dB cut-off at $\approx$8.2 GHz can be estimated at all the laser wavelengths (the dip at $\approx$ 1.5 GHz is a resonance in the detection chain that we did not manage to eliminate yet).

Figure~\ref{fig:3}(b) reports the BW, at a laser wavelength of 9.85 $\mu$m, for increasing RF input powers: 0, 3, 6 and 9 dBm. The measurements show that the modulator response is linear with the input power, and the BW keeps constant too. 
\begin{figure}[t]
\centering
\includegraphics[width=0.8\columnwidth]{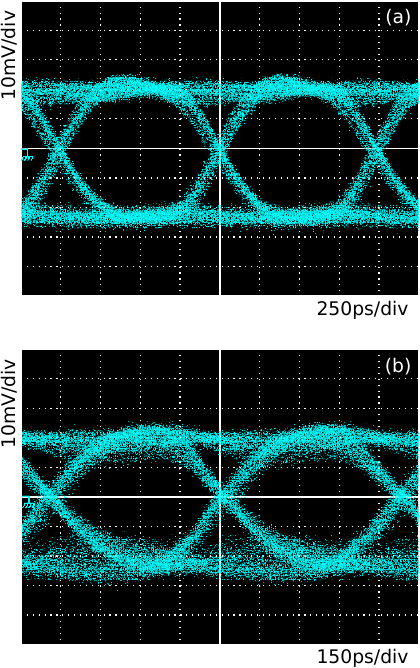}
\caption{\label{fig:4}Eye diagrams  at (a) 1Gbit/s and (b) 1.5Gbit/s (limited by VIGO MCT bandwidth). }
\end{figure}
The inset of Fig.~\ref{fig:3}(b) shows the dependence of the modulator response on the incident laser power, up to 24 mW. The response is linear up to 15 mW. A beginning of saturation appears between 15 and 24 mW.
This power is in accordance with the theory developed in Ref.~\cite{jeannin_unified_2021}, that suggests that low intensity saturation ($I_{sat}$) values, of the order of several kW/cm$^2$, are possible in these first-generation devices operating close to the onset of the strong light matter coupling regime.
We have recently experimentally measured such low values while developing semiconductor saturable absorption mirrors (SESAM) in the mid-IR~\cite{Jeannin2023}. Furthermore, evidence of responsivity decrease with increasing incident optical power in fast QWIPs has been recently attributed to optical saturation as well~\cite{Lin2023}.

A binary data modulation signal has been applied to display eye diagrams as a figure of merit of the ultra-fast modulators.
To this scope, we employed a commercial fast MCT detector (VIGO Systems) delivering more signal voltage at the output compared to our fast QWIP that - as seen previously - operates on the tail of its responsivity curve (see Fig. 2(c), gray line). 
The MCT detector used has a nominal bandwidth of 800 MHz, as detailed in Supplementary Material, that limits the observable data rate. 
Figure~\ref{fig:4} shows eye diagrams acquired with a pseudo-random binary sequence (PRBS) up to 1.5 Gbit/s.
The binary digit states of the numerical transmitted signal are defined respectively either with light transmission or light extinction. The high voltage level in the reported eye diagrams corresponds to the light transmission
The eye diagram exhibits a clear discrimination between low and high voltage levels and a wide time separation between the level transitions. The rise and fall times around 300 ps can be directly assigned to the 800 MHz-limited bandwidth of the detector characteristics. The modulator cut-off frequency effect being not visible on this display that confirms the cut-off frequency is well beyond the detector bandwidth.

In conclusion, we have exploited an electrically reconfigurable surface to demonstrate a free-space, room-temperature amplitude modulator for mid-infrared radiation with a -3dB cutoff at about 8.2 GHz.
The device operates in the 9.4-9.8 $\mu$m and 8.5-8.9 $\mu$m wavelength range, with RF input powers up to 9 dBm.
Future developments will focus on increasing the performance, in terms of modulation contrast, speed and spectral response in order to make these devices compatible with spectroscopy (short-term) and phase-modulation\cite{Chung2023} (long-term) applications.
In particular the speed can be increased by further reducing the device total surface (hence reducing the capacitance). This can be obtained by further separating the patch resonators. 
Finally, we have recently developed an optical analogous of this system, where such transition is \textit{optically} induced in order to implement SESAMs operating across the whole mid-IR~\cite{jeannin_unified_2021,Jeannin2023}.
\rosso{A cross-breeding of these two approaches is also possible, i.e. devices whose non-linear properties are tunable with an externally applied bias.}

%
\section*{Supplementary Material}
See supplementary material for (i) details of the RF modulation measurement setup; (ii) BW of the fast, commercial MCT detector; (iii) device fabrication; (iv) AR details; (v) S-parameters.

\section*{Data Availability Statement}
The data that support the findings of this study are available from the corresponding author upon reasonable request.

\begin{acknowledgments}
This work was partially supported by the \textit{Programme de Prematuration} of CNRS; the French National Research Agency: project SOLID (No. ANR-19-CE24-0003); 
and the European Union through  FET-Open Grant MIRBOSE (737017).\\
This work was done within the C2N micro nanotechnologies platforms and partly supported by the RENATECH network and the General Council of Essonne. 
We thank Farah Amar for help with design and RF test of the integrated coplanar lines; Alan Durnez, Francois Maillard and Xavier Lafosse for assistance in the C2N cleanroom; Cristiano Ciuti for granting us access to the Laboratoire Matériaux et Phénomènes Quantiques cleanroom (CNRS UMR 7162, Universit\'e Paris-Cit\'e); and Pascal Filloux for assistance in the Paris-Cit\'e  cleanroom. 
We thank Delphine Morini and Victor Turpaud for the loan of the pseudo-random bit sequence (PRBS) generator.

\end{acknowledgments}

\section*{References}
%

\end{document}